\documentclass[graybox]{svmult}

\usepackage{mathptmx}       
\usepackage{helvet}         
\usepackage{courier}        
\usepackage{type1cm}        
\usepackage{makeidx}         
\usepackage{graphicx}        
\usepackage{multicol}        
\usepackage[bottom]{footmisc}

\makeindex             

\def\bea{\begin{eqnarray}}
\def\eea{\end{eqnarray}}

\begin{document}

\title*{Localizing gravitational wave sources with optical telescopes and combining electromagnetic and gravitational wave data}
\author{Shaon Ghosh and Gijs Nelemans}
\institute{Shaon Ghosh \at Radboud University, Nijmegen and Nikhef, the Netherlands \email{shaon@astro.ru.nl}
\and Gijs Nelemans \at Radboud University, Nijmegen, and Nikhef, the Netherlands \email{nelemans@astro.ru.nl}}
%
%
\maketitle

\abstract{Neutron star binaries, which are among the most promising sources for the direct detection of gravitational waves (GW) by ground based detectors, are also potential electromagnetic (EM) emitters. Gravitational waves will provide a new window to observe these events and hopefully give us glimpses of new astrophysics. In this paper, we discuss how EM information of these events can considerably improve GW parameter estimation both in terms of accuracy and computational power requirement. And then in return how GW sky localization can help EM astronomers in follow-up studies of sources which did not yield any prompt emission. We discuss how both EM source information and GW source localization can be used in a framework of multi-messenger astronomy. We illustrate how the large error regions in GW sky localizations can be handled in conducting optical astronomy in the advance detector era. We show some preliminary results in the context of an array of optical telescopes called BlackGEM, dedicated for optical follow-up of GW triggers, that is being constructed in La Silla, Chile and is expected to operate concurrent to the advanced GW detectors.}

\section{Introduction}

The second generation GW detectors are scheduled to come online from 2015 and should be operating at their design sensitivities by the end of 2019 \cite{prospectsOfSkyloc}. A worldwide network of these advanced detectors will enable us in conducting GW astronomy for the first time in history. Compact binary coalescing (CBC) systems are the most promising candidates for detection of gravitational waves by these detectors as their GW frequency sweep through the sensitivity band towards merger. Binary neutron stars (BNS) and neutron star black hole binaries (NSBH) are also prospective progenitors of short duration gamma ray bursts (GRB). Current models of GRB mechanisms \cite{Nakar:2007yr, Piran:1999kx} predicts that the prompt emissions are launched just before the merger of the compact binary. This presents to us a unique opportunity for joint EM-GW astronomy. EM astronomy could complement GW astronomy in a number of ways. On one hand, gravitational wave sky-localization for the advanced ground based detectors could have large uncertainties, typically $\sim 10-100$ square degrees \cite{TriangulationSteve}. Typical sky-localization of arc-second accuracy could be achieved in conventional EM astronomy. On the other hand, gravitational waves observation allows us to make an independent and direct measurement of masses and luminosity distances of these sources. Evidently, a tremendous amount of scientific merit lies in combining EM and GW observations. Implementation of EM information for detection of gravitational waves is already an active area of research. LIGO Scientific collaboration (LSC) and Virgo Scientific collaboration have been conducting search for gravitational waves from progenitors of short duration gamma ray bursts (sGRB) by following up on GCN (Gamma-ray Coordination Network) triggers \cite{s5grb, s6grb}. In the context of GW parameter estimation, the use of EM source information gets an even richer dimension. A non-spinning compact binary system can be characterized by nine parameters, the luminosity distance $d_L$, the binary component masses $m_1$ and $m_2$, the sky position of the binary $(\alpha, \delta)$, the inclination angle $\iota$ of the axis of the binary w.r.t the observer's line of sight, the polarization angle $\psi$ of the binary orbit, the time of arrival of the signal $t_a$ and the coalescence phase of the binary $\delta_0$. EM information can fix (or constrain) a subset of these parameters reducing the dimensionality of the posterior probability distribution of \cite{RoeverMCMC, sweta} which directly translate into considerable reduction in computational power requirements for the estimation of the source parameters. In this paper, we first show in Sec. \ref{sec:EMPriors} how the EM information can improve the estimation of the parameters. We examine how different types of EM information can help in the estimation of inclination angle of binary and luminosity distance to the source. Then in Sec. \ref{sec:GWskyLoc} we explore how GW parameter estimations can in turn aid EM astronomy. Here we focus entirely on sky localization of gravitational wave sources. We discuss how to generate telescope pointing for optical telescope from error regions of the sky localization. In that context, we acknowledge the efforts towards construction of a dedicated GW follow-up facility.

\section{Gravitational wave parameter estimation in presence of electromagnetic information}
\label{sec:EMPriors}

Most likely sources of electromagnetic information for gravitational wave parameter estimation could be in the form of prompt emissions like short duration gamma ray bursts, or from associated afterglows and kilonovae. A fully coherent parameter estimation incorporates a Bayesian Markov chain Monte Carlo technique which is computationally expensive. Thus it is extremely important that we choose a representative system. For this study, we injected inspiral gravitational wave signals from the TaylorT4 waveform family with order 3.5PN \cite{TaylorT4} in initial LIGO colored Gaussian noise. The source was chosen to be a neutron star - black hole binary (NSBH), with a $1.4 M_{\odot}$ neutron star and a $10 M_{\odot}$ black hole, inclined at angle of $10^{\circ}$ w.r.t the observers line of sight, located at a distance of $\sim 30$ Mpc from earth.\\ The choice of a NSBH system is motivated by the fact that a fair amount of research is ongoing on binary neutron star (BNS) systems while parameter estimation studies of NSBH systems are virtually non-existent. The choice of the inclination angle was motivated by the fact that even though there exists some uncertainties regarding the bounds on short GRB opening angles, recent studies \cite{grbOpeningAngle} estimated the median opening angles of short GRBs to be $\approx 10^{\circ}$. 

\begin{figure}[tbh]
\centering
\includegraphics[width=5cm]{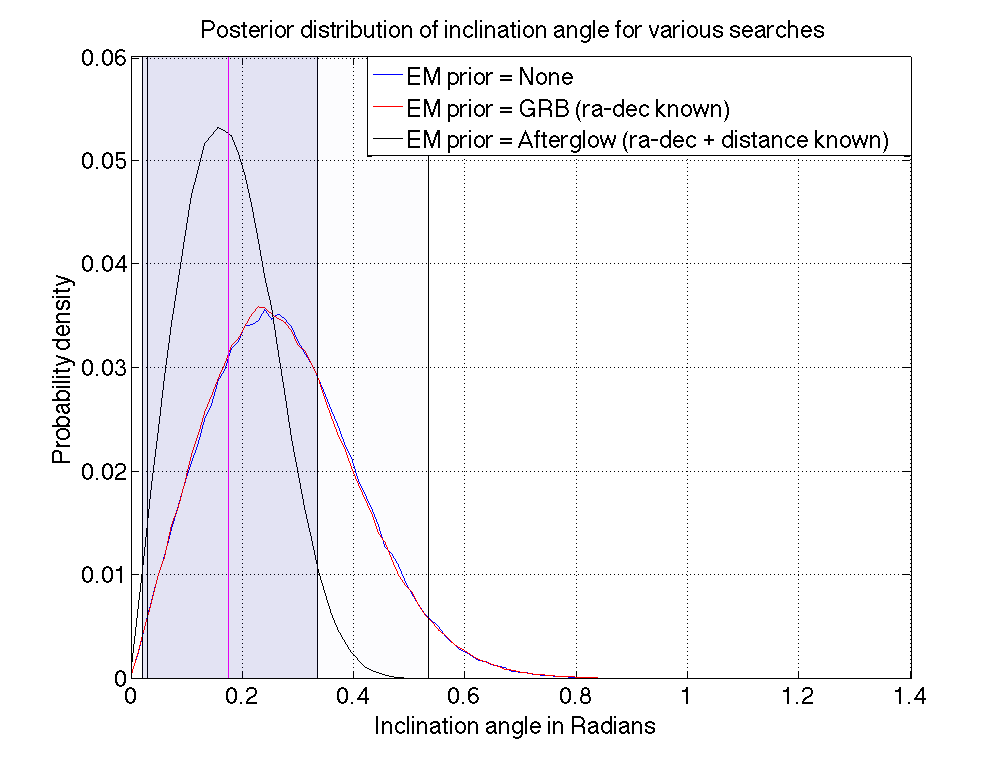}
\includegraphics[width=5cm]{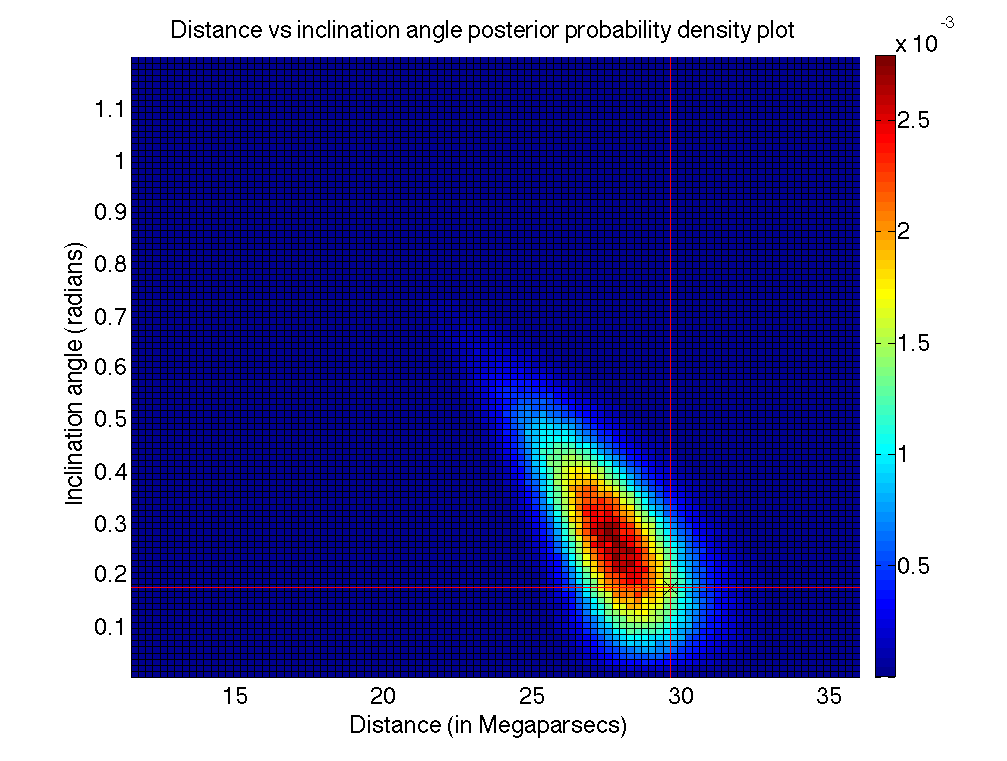}
\includegraphics[width=5cm]{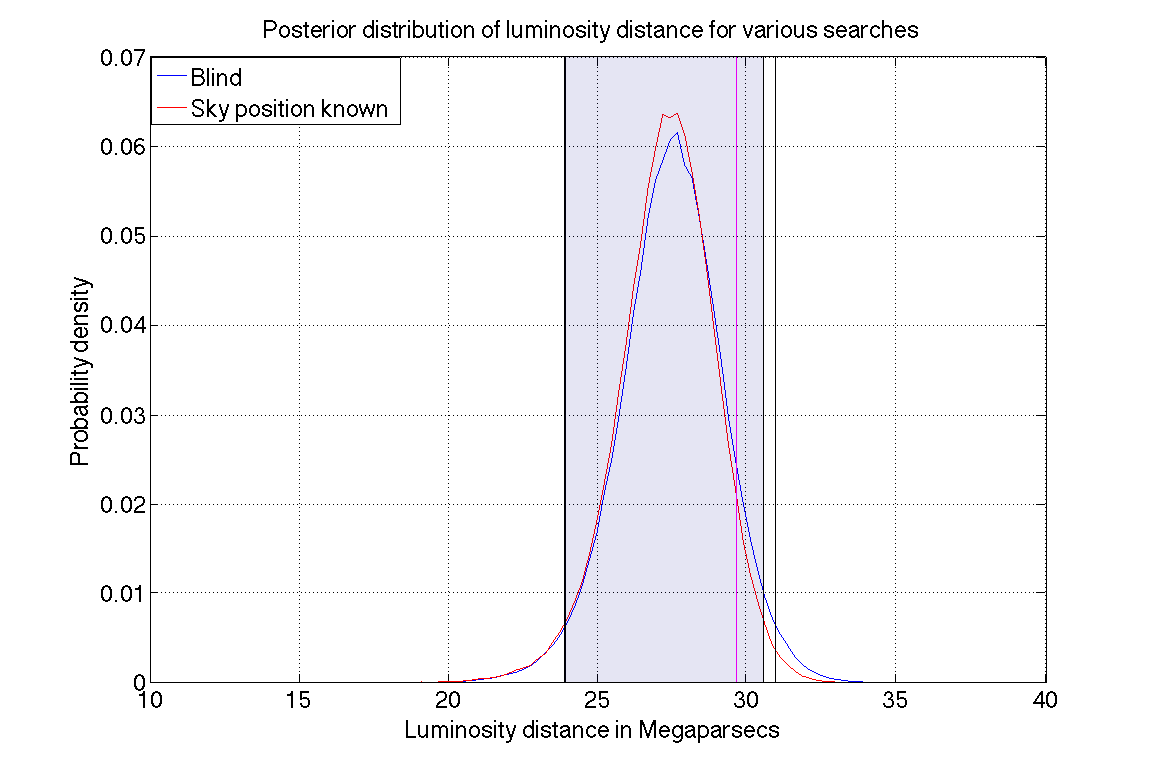}
\includegraphics[width=5cm]{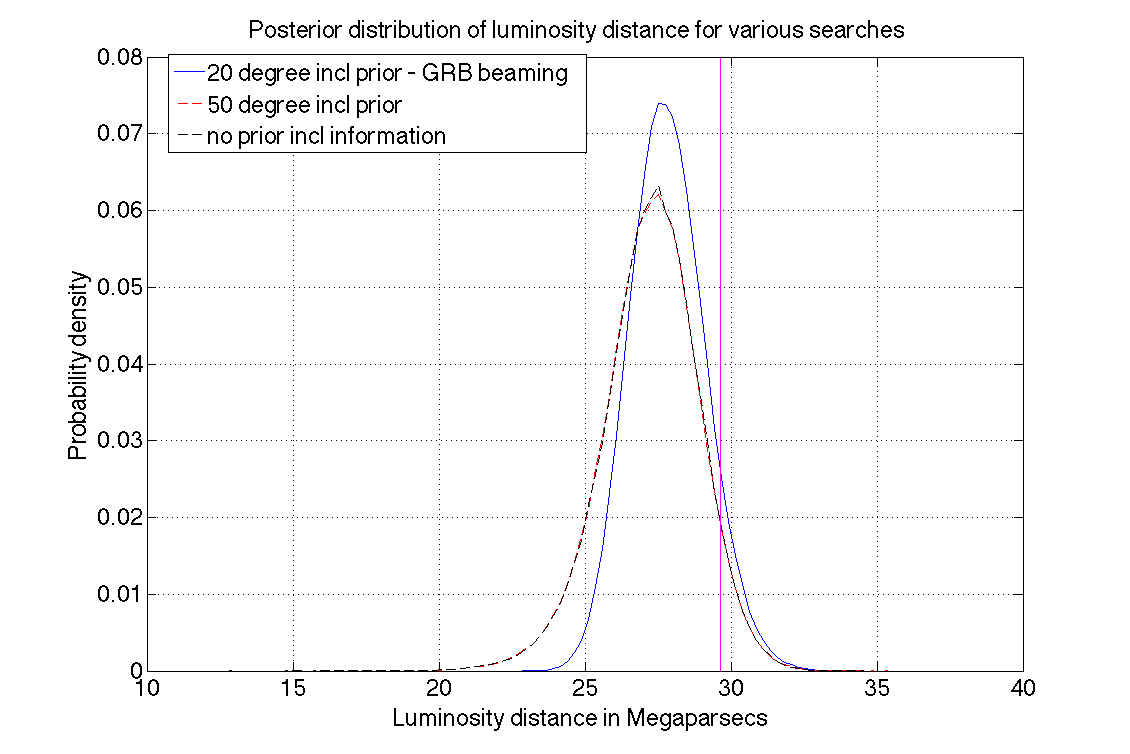}
\caption{{\em Top left}: Estimation of inclination angle for the three classes of studies, blind, sky position known and sky position and distance known. The shaded region denotes the $2\sigma$ spread of the distribution and the vertical magenta line denotes the actual injected inclination angle. Note that in the presence of distance information the inclination angle estimation gets better. {\em Top right}: The $2D$ posterior probability distribution plot for the inclination angle and distance shows the strong correlation between the two parameters.  {\em Bottom left} No measurable improvement in estimation of luminosity distance upon utilizing sky position information. {\em Bottom right}: Improvement in estimation of distance in the presence of inclination angle constraints. }
\label{fig:iotaCorrel}
\end{figure}

For the comparative analysis of how EM information can help GW parameter estimation we conducted three classes of studies. The first study constitutes the control of our experiment in the form of `blind' parameter estimation where we have no EM information available at our disposal. In the second class of analysis, we used the sky position of the source information, an astrophysical instance of which would be the case where a GRB was detected and located by the gamma ray observatory. And finally the third class of analysis involved the use of sky position and distance information. This corresponds to the case where a GRB and a subsequent afterglow associated with the same were discovered thus conveying to us the sky position and the distance information in the form of redshift. We found that not all parameters displayed improvement in estimation upon use of EM information. However, at the same time we note that certain parameters that are strongly correlated with another parameter, showed considerable improvement in its estimation upon using the correlated parameter EM information. As an example, we show the estimation of inclination angle in the presence of various EM information in figure \ref{fig:iotaCorrel}.
We note from this study that an sky position information does not improve inclination angle estimation. However, knowledge of distance greatly improves the precision and accuracy of the estimation as shown in figure \ref{fig:iotaCorrel} (top left). This is due to the well known strong correlation between the inclination angle and distance that we have shown in figure \ref{fig:iotaCorrel} (top right). Though we have seen in figure \ref{fig:iotaCorrel} (bottom left) that the knowledge of sky position itself do not provide any improvement in the estimation of distance, however one should note that if the knowledge of the sky position came from the detection of a GRB, then one might put constrains on the inclination angle of the source. This stems from the fact that for a GRB to be observed it would indicate that the source might be more favorably oriented to the observer. We show that as a result of constraining of inclination angle to $20^{\circ}$, the estimation of the distance in figure \ref{fig:iotaCorrel} (bottom right) improves compared to unconstrained distance estimation. 

\begin{figure}[tbh]
\centering
\includegraphics[width=5cm]{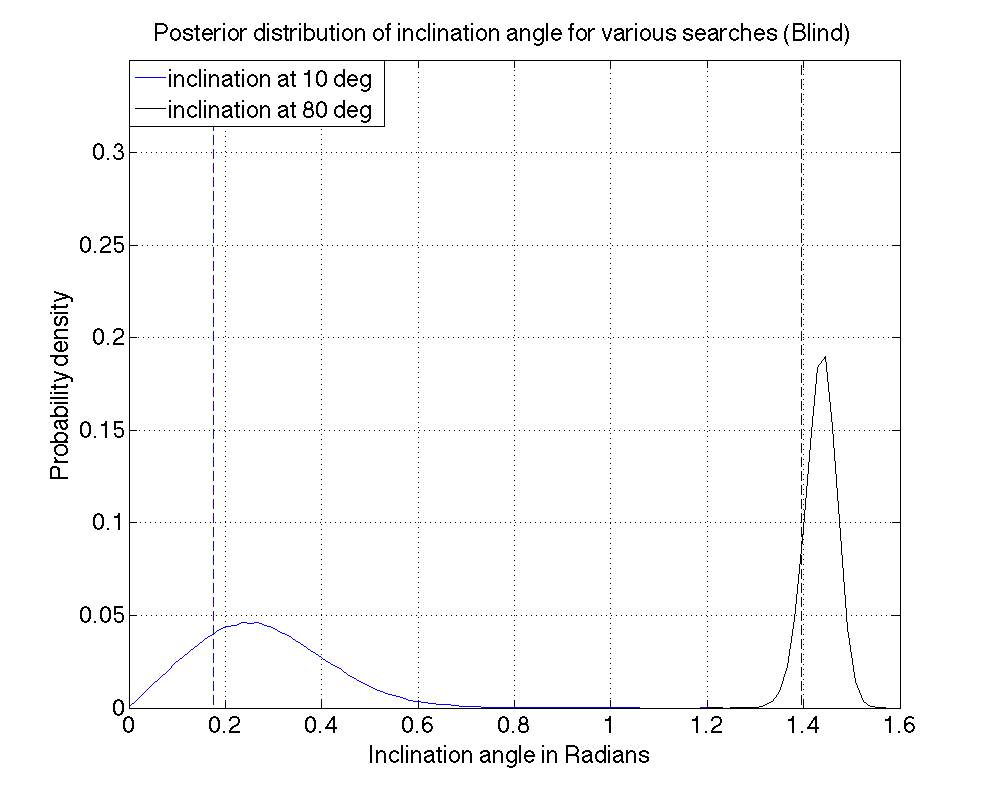}
\includegraphics[width=5cm]{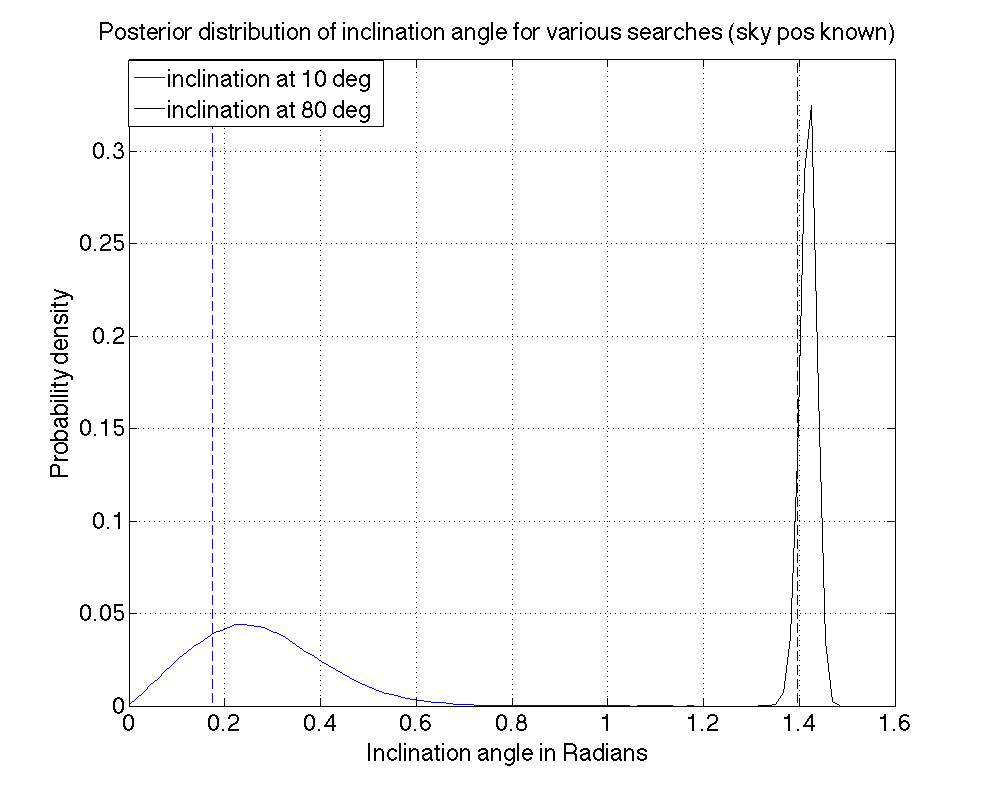}
\caption{{\em Left}: Inclination angle posterior probability distributions showing dependence on the inclination angle of the injected source. We observe that the edge-on source has the smallest spread. Note that higher PDFs are narrower. {\em Right}: Similar study conducted for systems for which the sky position is known from EM observations. }
\label{fig:edgeOnvsFaceOn}
\end{figure}

Studies of edge-on systems revealed some interesting features. Edge-on systems are different from face-on systems owing to the lack of the cross-polarization. Parameter estimation algorithm very quickly realizes its absence in the date and consequently estimates the inclination angle very precisely. This is observed in figure \ref{fig:edgeOnvsFaceOn} (left), where we have compared the estimation of the inclination angle for an edge-on source with face-on oriented source. We conducted a similar study for sources with EM information in the form of sky position in figure. \ref{fig:edgeOnvsFaceOn} (right). While sky position and inclination do not have strong correlation and therefore not much benefit is expected in the inclination angle estimation from sky position information, as we have seen in figure \ref{fig:iotaCorrel} (bottom left) red and blue plots, the situation is very different for edge-on systems. In figure \ref{fig:edgeOnvsFaceOn} we compare the results of inclination angle estimation of edge-on and face-on systems in the absence (left) and presence (right) of sky position information. It is clear that the presence of sky position information strongly aids the inclination angle estimation for the edge-on configuration, unlike the previously shown face-one case. This can be explained as follows. The strain at a detector due to a particular polarization does not depend only on the inclination angle but also depends on the sky position. 
\bea
h(t) = F_+(\theta, \phi, \psi)h_+(r, \iota, m_1, m_2, \delta_c, t_c) + F_{\times}(\theta, \phi, \psi)h_{\times}(r, \iota, m_1, m_2, \delta_0, t_c)\,,
\eea
where the $F_+(\theta, \phi, \psi)$ and $F_{\times}(\theta, \phi, \psi)$ are the antenna pattern functions and $h_+$ and $h_{\times}$ are the two gravitational wave polarization components. In absence of sky position information, we essentially have the freedom to change the antenna pattern functions during the estimation of the inclination angle and a wider posterior probability distribution could be accommodated due to this freedom. Thus, when the sky position information is used, we are likely to get narrower estimate of the inclination angle.

\section{Gravitational wave sky localization for electromagnetic follow-up studies}
\label{sec:GWskyLoc}

Past observations have not yielded any detection of short duration GRB within the typical advanced LIGO range $\sim 300$ Mpc, for which we have redshift information  \cite{Nakar:2007yr}. This could partly be explained by the lower event rate compared to their longer duration counterparts and strong beaming mechanism. Gravitational wave radiation on the other hand is expected to be more isotropic. Thus, gravitational wave source localization could act as a pointer for optical follow-up studies of short GRBs. Short GRB afterglows progressively gets more isotropic as we go from lower wavelength X-ray to higher wavelength optical band. An optical afterglow associated with a prompt emission is expected to be visible by ground based telescopes for typically $\sim 1$ day. Kilonovae produced by r-process decay of neutron rich radioactive elements created due the disruption of the neutron star in a neutron star binary is expected to be highly isotropic. The emission from these events could predominantly be in the near-infrared regime \cite{kilonova, kilonovaOpacity}. These events might or might not accompany a short GRB, thus have the potential of being an optical/near-infrared transient candidate associated with a class of compact binary mergers independent of whether or not they produce a prompt emission.\\ In order to meaningfully follow-up EM counterparts of gravitational wave triggers it is of paramount importance to identify these events very quickly \cite{svdPaper}, and then localize them with very low latency.  Typical time scales of a fully Bayesian Markov chain Monte Carlo (MCMC) study is of the order of a day, hence not useful for the purpose of optical followup. In order to address this issue, a rapid sky  localization technique, known as BAYESTAR, has been developed that is capable of localizing gravitational wave events in the sky within minutes of receiving the trigger using techniques based on timing, phase and amplitude consistency \cite{first2years}. We generate gravitational wave error regions for these events which can then be followed-up by optical and infrared telescopes around the world. However, it is important to recognize that in the first year of advanced detectors only the LIGO detectors will be operating. A less sensitive Virgo detector will be coming online a year after. The error regions in these first two years are going to be elongated shaped and large with a median of $500$ ($200$) square degrees with $90\%$ confidence level in first year (second year) of opearation respectively \cite{first2years}. Under such circumstances, it is important to have telescopes that can cover such error regions within the optical transient timescales. We are currently constructing an array of telescopes in La Silla, Chile, called the BlackGEM that will initially (2016) have four telescopes, each with a field of view of $2.7$ square degrees, that can potentially reach up to 23rd magnitude in 5 minutes of integration time. The full BlackGEM array of 15 telescopes is aimed to become operational in 2018. The multiple telescopes in the array can adjust their configurations to morph the field of view of the array to the shape of the error region. We created telescope pointing for BlackGEM using a number of sky-localizations received from the low latency timing pipeline to see how well BlackGEM is able to cover them. In our preliminary studies, we used 10 injections that were localized using the low latency pipeline. Half of the injections were found as doubles and half as triples. We found that a four-telescope array will be pressed hard to cover error region for two detectors. Typical error region from LIGO only sky-localizations required $> 50$ pointing from BlackGEM. With each pointing requiring 5 minutes of integration time, it will require many hours to cover the error regions completely. Moreover, we found that the error regions can span across both the hemispheres indicating it may not always be feasible to cover the entire error region by one observatory. 
\begin{figure}[tbh]
\centering
\includegraphics[width=10.0cm]{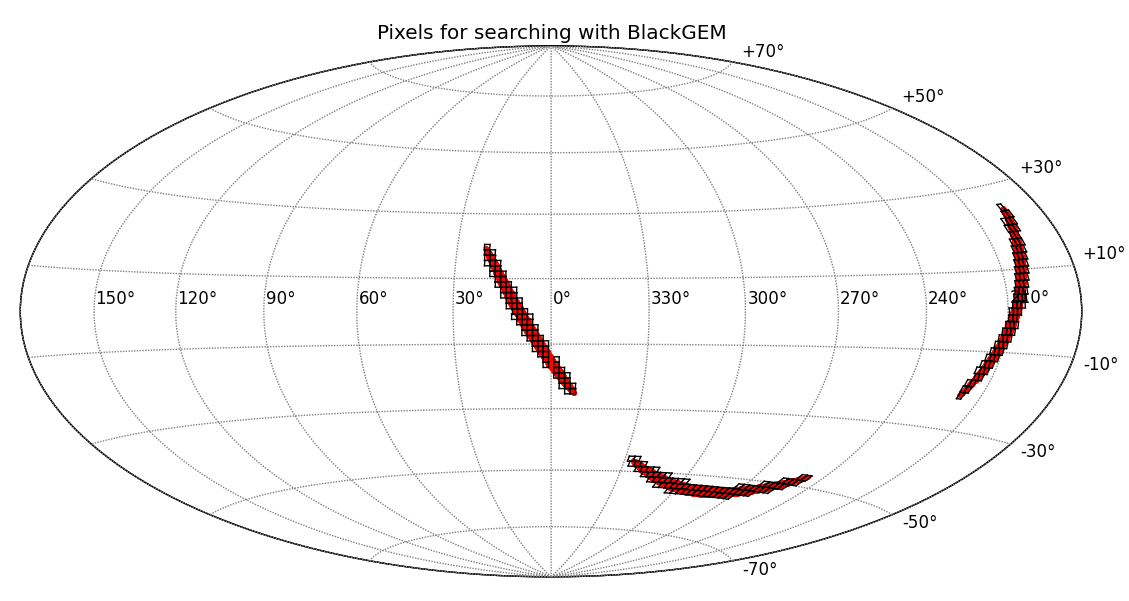}
\caption{BlackGEM pointing of $2\sigma$ error region for a double coincident detection in LIGO interferometer. Note that the long arcs span across both the hemisphere. One telescope might not be able to fully follow-up the entire error regions for all cases. }
\end{figure}
The situation gets better when Virgo comes online. With the error regions generally more localized and located within a single hemisphere, we found that we are in a position to cover the desired region within $6-20$ BlackGEM pointing for sources lasting in time scales of hours. With more number of telescopes in the array, it will be realistically possible to cover most of the error regions from three detector localizations, and we expect that the full array of 15 telescopes should be able to cover a large fraction of two detector error regions.

\begin{figure}[tbh]
\centering
\includegraphics[width=10.0cm]{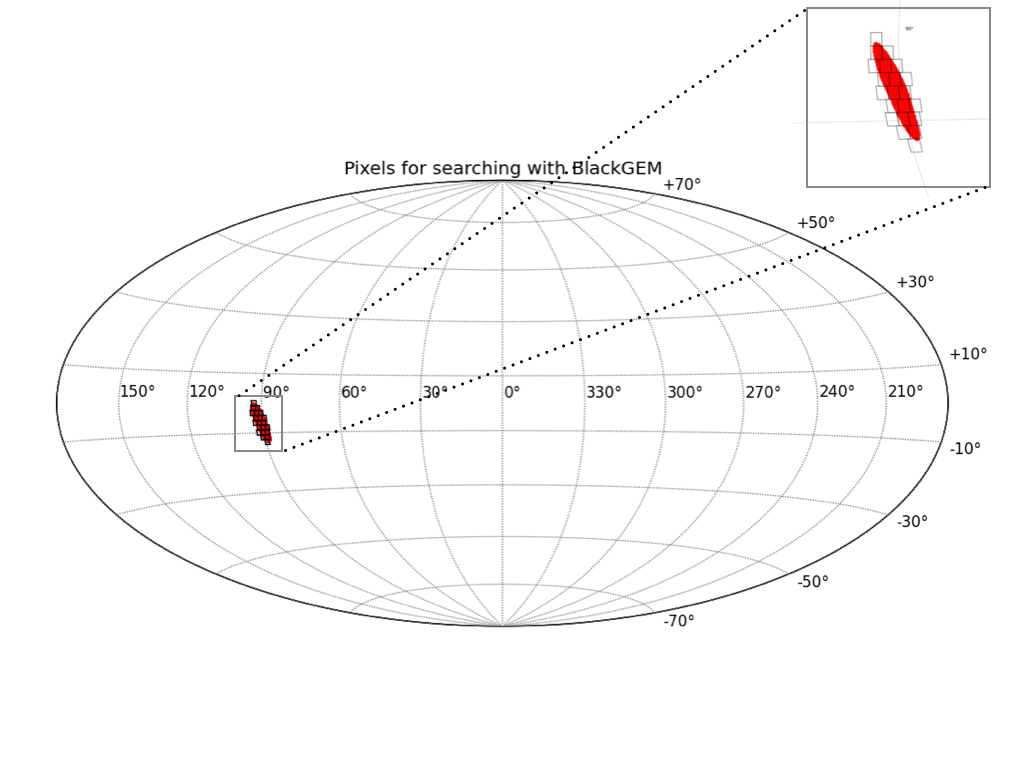}
\caption{BlackGEM pointing of $2\sigma$ error region for a triple coincident detection in LIGO-Virgo interferometers. The error region in this case can be more realistically followed-up with BlackGEM.}
\end{figure}

\section{Discussion}
In this work, we explored various avenues of improving GW parameter estimation of NSBH systems using EM information from GRB and/or associated afterglows and Kilovovae. Our results show that EM information in the form of redshift from afterglows could be very useful for GW parameter estimation due to strong $d_L-\iota$ correlation. We also examined the improvement upon constraining inclination angle to $20^{\circ}$ for known GRB triggers. We then explored the methods of conducting optical follow-up of GW triggers and presented some preliminary results of this study for the BlackGEM array of telescopes. We observed that it will be difficult to cover the full LIGO $2\sigma$ error regions in the first year when the Virgo detector is not operating. With the second year of operation of LIGO as the sensitivities improve, as Virgo comes online and as the full BlackGEM array is constructed, we would be seeing realistic possibilities of covering the GW error regions and possibly start making concurrent observations of EM counterparts to GW events.

\section{Acknowledgement}

We would like to acknowledge Steven Bloemen and Paul Groot for providing us the sky pixel data of BlackGEM that was utilized to construct the pointing. SG is thankful to Marc van der Sluys for the discussions on the techniques of parameter estimation and MCMC. We would also like to thank Larry Price and Leo Singer for their assistance regarding the use of the low latency sky localization pipeline and constructive feedback regarding the work.

\bibliographystyle{ieeetr}
\bibliography{References}
\end{document}